\pdfoutput=1

\documentclass[11pt,twoside,a4paper,cmspaper,final,collab]{cms-tdr}
\def\svnVersion{218306}\def\cmsCernNo{2013-093}\def\cmsCernDate{\today}\def\cmsMessage{Published in Physical Review Letters as \href{http://dx.doi.org/10.1103/PhysRevLett.111.221801}{\doi{10.1103/PhysRevLett.111.221801.}}}

\begin{document}\cmsNoteHeader{SUS-13-003}

\hyphenation{had-ron-i-za-tion}
\hyphenation{cal-or-i-me-ter}
\hyphenation{de-vices}

\RCS$Revision: 213980 $
\RCS$HeadURL: svn+ssh://svn.cern.ch/reps/tdr2/papers/SUS-13-003/trunk/SUS-13-003.tex $
\RCS$Id: SUS-13-003.tex 213980 2013-10-28 13:56:45Z mwalker $
\newcommand{\fullLumi}{19.5\fbinv\xspace}
\newcommand{\procLumi}{19.5\fbinv\xspace}
\newcommand{\ST}{\ensuremath{S_\mathrm{T}}\xspace}
\newcommand {\abseta}{\ensuremath{\abs{\eta}}\xspace}
\providecommand{\tauh}{\ensuremath{\tau_\mathrm{h}}\xspace}

\newlength\cmsFigWidth
\ifthenelse{\boolean{cms@external}}{\setlength\cmsFigWidth{0.85\columnwidth}}{\setlength\cmsFigWidth{0.4\textwidth}}
\ifthenelse{\boolean{cms@external}}{\providecommand{\cmsLeft}{top}}{\providecommand{\cmsLeft}{left}}
\ifthenelse{\boolean{cms@external}}{\providecommand{\cmsRight}{bottom}}{\providecommand{\cmsRight}{right}}
\ifthenelse{\boolean{cms@external}}{\providecommand{\suppMaterial}{the supplemental material [URL will be inserted by publisher]\xspace}} {\providecommand{\suppMaterial}{Appendix~\ref{app:suppMat}}\xspace}
\cmsNoteHeader{SUS-13-003} 
\title{Search for top squarks in \texorpdfstring{$R$}{R}-parity-violating supersymmetry using three or more leptons and \texorpdfstring{$\cPqb$}{b}-tagged jets}

\date{\today}

\abstract{
A search for anomalous production of events with three or more isolated leptons and bottom-quark jets produced in $\Pp\Pp$ collisions at $\sqrt{s} = 8$\TeV is presented. The analysis is based on a data sample corresponding to an integrated luminosity of 19.5\fbinv collected by the CMS experiment at the LHC in 2012. No excess above the standard model expectations is observed. The results are interpreted in the context of supersymmetric models with signatures that have low missing transverse energy arising from light top-squark pair-production with $R$-parity-violating decays of the lightest supersymmetric particle. In two models with different $R$-parity-violating couplings, top-squarks are excluded below masses of 1020\GeV and 820\GeV when the lightest supersymmetric particle has a mass of 200\GeV.
}

\hypersetup{%
pdfauthor={CMS Collaboration},%
pdftitle={Search for top squarks in R-parity-violating supersymmetry using three or more leptons and b-tagged jets},%
pdfsubject={CMS},%
pdfkeywords={CMS, physics, R-parity violation, squarks}}

\maketitle 
Supersymmetric (SUSY) extensions of the standard model (SM) solve the hierarchy problem while unifying particle interactions~\cite{Nilles:1983ge,Haber:1984rc}. Among SUSY models, ``natural" supersymmetry refers to those characterized by small fine tuning needed to describe particle spectra. It requires top squarks (stops), the top-quark superpartners, to be lighter than about 1\TeV. These models have received substantial interest in light of the discovery of a Higgs boson with mass near 125 \GeV~\cite{Chatrchyan:2012gu,Aad:2012gk} because the stop should be the superpartner most strongly coupled to the Higgs.

Natural models feature pair production of stops that decay to a number of final states. To fully test supersymmetric naturalness, searches for all possible decay chains should be carried out. These can be broadly categorized as $R$-parity conserving (RPC) or violating (RPV)~\cite{Farrar:1978xj}, where $R$-parity is defined by $R = (-1)^{3B+L+2s}$, with $B$ and $L$ the baryon and lepton numbers, and $s$ the particle spin. All SM particle fields have $R = +1$ while all superpartner fields have $R = -1$. When $R$-parity is conserved, superpartners are produced in pairs, the lightest superpartner (LSP) is stable and a dark-matter candidate, and proton stability is ensured. Most recent searches for naturalness have focused on RPC models~\cite{Chatrchyan:2013wxa,Chatrchyan:2013xna,Aad:2013ija}.

Supersymmetric models with RPV interactions violate either $B$ or $L$ but can avoid proton decay limits~\cite{Barbier:2004ez,PhysRevD.86.010001}. The superpotential $W_\mathrm{RPV}$ includes three trilinear terms parametrized by the Yukawa couplings $\Lam_{ijk}$, $\Lamp_{ijk}$, and
$\Lampp_{ijk}$:
\begin{equation}
\label{eqn:wrpv}
W_\mathrm{RPV} =
\frac{1}{2}\, \lambda_{ijk}L_iL_j\overline{E}_k+
\lambda^\prime_{ijk}L_iQ_j\overline{D}_k\,+\,
\frac{1}{2}\, \lambda^{\prime \prime}_{ijk}\overline{U}_i\overline{D}_j\overline{D}_k,
\end{equation}
where $i,j,$ and $k$ are generation indices; $L$ and $Q$ are the $SU(2)_L$ doublet superfields of the lepton and quark; and the $\overline{E}$, $\overline{D}$, and $\overline{U}$ are the $SU(2)_L$ singlet superfields of the charged lepton, down-like quark, and up-like quark.  The third term violates baryon number conservation, while the first two terms violate lepton number conservation. These terms do not preclude a natural hierarchy~\cite{Evans:2012bf}.

The RPV interactions allow for single production of SUSY particles (sparticles) and for sparticle decay into SM only particles. The latter is explored in this Letter. Prior searches for RPV interactions in multilepton final states include those at LEP~\cite{Heister:2002jc,Abdallah:2003xc,Achard:2001ek}, the Tevatron~\cite{Abazov:2006nw,Abulencia:2007mp}, at HERA~\cite{HERA-LQD,zeus:lqdstop}, and at the Large Hadron Collider (LHC)~\cite{RA7PLB,Chatrchyan:2012mea,ATLAS:2012kr}.

Because the LSP is unstable in RPV models, a common experimental strategy of SUSY searches --- selecting events with large missing transverse energy ($\MET$) --- is not effective~\cite{Barbier:2004ez}. Instead, we use $\ST$, the scalar sum of $\MET$ and the transverse energy of jets and charged leptons, to differentiate between signal and standard model backgrounds.

In this Letter we present the result of a search for pair production of top squarks with RPV decays of the lightest sparticle, using multilepton events and bottom-tagged ($\cPqb$-tagged) jets. The dataset used here corresponds to an integrated luminosity of $\procLumi$, recorded in 2012 with the CMS detector at the LHC in proton-proton collisions at a center-of-mass energy of 8\TeV.

The coordinate system in CMS is right-handed, with the origin at the nominal interaction point. Pseudorapidity is given by $\eta \equiv -\ln[\tan(\theta/2)]$, where the polar angle $\theta$ is defined with respect to the counterclockwise beam direction. The azimuthal angle $\phi$ is measured relative to the direction to the center of the LHC ring.

The CMS detector~\cite{Chatrchyan:2008zzk} has cylindrical symmetry around the $\Pp\Pp$ beam axis with tracking and muon detectors covering the pseudorapidity range $\abseta<2.4$.  The tracking system measures the trajectory and momentum of charged particles and consists of multilayered silicon pixel and strip detectors in a 3.8\unit{T} solenoidal magnetic field. Particle energies are measured with concentric electromagnetic  and hadron calorimeters, which cover $\abseta < 3.0$ and $\abseta < 5.0$, respectively. Muon detectors consisting of wire chambers are embedded in the steel return yoke outside the solenoid. The trigger thresholds in a two-level trigger system are tuned to accept
a few hundred data events per second from the $\Pp\Pp$ interactions.

We select events with three or more leptons (including tau leptons) that are accepted by a trigger requiring two light leptons, which may be electrons or muons. Any opposite-sign, same-flavor (OSSF) pair of electrons or muons must have an invariant mass $m_{\ell \ell} > 12$\GeV, removing low-mass bound states and $\gamma^{*}\rightarrow \ell^{+} \ell^{-}$ production.

Electrons and muons are reconstructed using the tracker, calorimeter, and muon systems. Details of reconstruction and identification can be found in Ref.~\cite{EGM-10-004} for electrons and in Ref.~\cite{Chatrchyan:2012xi} for muons. We require that at least one electron or muon in each event have transverse momentum of $\pt > 20$\GeV. Additional electrons and muons must have $p_{\rm T} > 10$\GeV and all of them must be within $|\eta| < 2.4$.

The majority of hadronic decays of tau leptons ($\tauh$) yield either a single charged track (one-prong) or three charged tracks (three-prong), occasionally with additional electromagnetic energy from neutral pion decays. We use one- and three-prong $\tauh$ candidates that have $\pt > 20$\GeV, reconstructed with the ``hadron plus strips'' method~\cite{HPStaus}. Leptonically decaying taus are included with other electrons and muons.

To ensure that electrons, muons, and $\tauh$ candidates are isolated, we use a particle-flow (PF) algorithm~\cite{PFT-10-004,PFT-08-001} to identify the source of transverse energy deposits in the trackers and calorimeters. We then sum the energy deposits in a cone of radius 0.3 in $\DR = \sqrt{(\Delta \eta)^2 + (\Delta \phi)^2}$ around the candidate and subtract the lepton $\pt$ to calculate $E_\text{T,cone}$. We remove energy from additional proton-proton collisions that occur simultaneously by subtracting a per-event correction~\cite{PFT-10-002,EGM-10-004}. For electrons and muons, we divide $E_\text{T,cone}$ by the lepton $\pt$ to find the relative isolation $I_\text{rel} = E_\text{T,cone}/\pt $, which has to be less than 0.15. We require $E_\text{cone} < 2$\GeV for $\tauh$ candidates.

We use jets reconstructed from PF candidates~\cite{PFT-10-002} using the anti-\kt algorithm~\cite{Cacciari:2008gp} with a distance parameter of 0.5, that have $\abseta < 2.5$ and $\pt > 30$\GeV.  Jets are required to be a distance $\Delta R > 0.3$ away from any isolated electron, muon, or $\tauh$ candidate. To determine if the jet originated from a bottom quark, we use the combined secondary-vertex algorithm, which calculates a likelihood discriminant using the track impact parameter and secondary-vertex information. This discrimination selects heavy-flavor jets with an efficiency of 70\% and suppresses light-flavor jets with a misidentification probability of 1.5\%~\cite{Chatrchyan:2012jua}.

Monte Carlo (MC) simulations are used to estimate some of the SM backgrounds and to understand the efficiency and acceptance of the signal models. The SM background samples are generated using \MADGRAPH~\cite{Maltoni:2002qb} with parton showering and fragmentation modeled using \PYTHIA (version 6.420)~\cite{Sjostrand:2007gs} and passed through a \GEANTfour-based~\cite{Agostinelli:2002hh} representation of the CMS detector. Signal samples~\cite{Evans:2012bf} are generated with \MADGRAPH and \PYTHIA and passed through the CMS fast-simulation package~\cite{Orbaker:2010zz}. Next-to-leading- and next-to-leading-log cross sections and their uncertainties for the SUSY signal processes are from the LHC SUSY cross sections working group~\cite{Beenakker:1999ch,Kulesza:2008jb,Kulesza:2009kq,Beenakker:2009ha,Beenakker:2011fu}.

Multilepton signals have two main sources of backgrounds, the first arising from processes that produce genuine multilepton events. The most significant examples are $\PW\cPZ$ and $\cPZ\cPZ$ production, but rare processes such as \ttbar\Wpm and \ttbar\cPZ\  also contribute.  We assess the contribution from these processes using samples simulated by \MADGRAPH. Samples simulating $\PW\cPZ$ and $\cPZ\cPZ$ have been validated in control regions in data. For the rarer background processes, we rely solely on simulation.

The second source originates from objects that are misclassified as prompt, isolated leptons, but are actually hadrons, leptons from a hadron decay, etc. Misidentified leptons are classified in three categories: misidentified light leptons (electrons and muons), misidentified $\tauh$ leptons, and light leptons originating from asymmetric internal conversions. The methods used in this paper are described in more detail in Ref.~\cite{Chatrchyan:2012mea}

We estimate the contribution of misidentified light leptons by measuring the number of isolated tracks and applying a scale factor between isolated leptons and isolated tracks. These scale factors are measured in control regions that contain leptonically decaying \cPZ-bosons and a third, isolated track, as well as in control regions with opposite-sign, opposite-flavor leptons, which are \ttbar-dominated. The scale factor is then the probability for the third track to pass the lepton identification criteria. We find the scale factors to be ($0.9 \pm 0.2$)\% for electrons and ($0.7 \pm 0.2$)\% for muons. The scale factors are applied to the sideband region with two light leptons and an isolated track. The scale factors depend on the heavy-flavor content in the different signal regions. We parameterize this dependance as a function of the impact parameter distribution of non-isolated tracks. The $\ttbar$ contribution is taken from simulation.

The $\tauh$ misidentification rate is measured in jet-dominated data by comparing the number of $\tauh$ candidates in the signal region defined by $E_{\rm cone} < 2$\GeV to the number of non-isolated $\tauh$ candidates, which have $6 < E_{\rm cone} < 15$\GeV. We measure the average misidentification rate as 15\% with a systematic uncertainty of 30\% based on the variation in different control samples. We apply this scale factor to the sideband region with two light leptons and one non-isolated $\tauh$ candidate.

Another source of background leptons is internal conversions, where a virtual photon decays to a dilepton pair. These conversions produce muons almost as often as electrons, and have been discussed in detail elsewhere~\cite{Chatrchyan:2012mea}. We measure the conversion factors of photons to light leptons in a control region (low \MET and low hadronic activity). The ratio of the number of $\ell^+\ell^-\ell^{\pm}$ candidates to the number of $\ell^+\ell^-\gamma$ candidates in the Z boson decays defines the conversion factor, which is  $ 2.1\%\pm1.0\%$ ($0.5\%\pm0.3\%$) for electrons (muons).

A systematic uncertainty of 4.4\% in the normalization of the simulated samples accounts for imperfect knowledge of the integrated luminosity of the data sample~\cite{CMS-PAS-LUM-12-001}. Signal cross sections have uncertainties from 15\% to 51\% in for stop masses between 250\GeV and  1.5\TeV, which come from the parton distribution function uncertainties and the renormalization and factorization scale uncertainties~\cite{Kramer:2012bx}. We scale the $\PW\cPZ$ and $\cPZ\cPZ$ simulation samples to match data in control regions. The overall systematic uncertainty on $\PW\cPZ$ and $\cPZ\cPZ$ contributions to the signal regions varies between 15\% and 30\% depending on the kinematics, and is the combination of the normalization uncertainties with resolution uncertainties. Muon identification efficiency uncertainty is 11\% at muon $\pt$ of 10\GeV and 0.2\% at 100\GeV. For electrons the uncertainties are 14\% at 10\GeV and 0.6\% at 100\GeV. The uncertainty on the efficiency of the bottom-quark tagger is 6\%. The uncertainty on the \MET resolution contributes a 4\% uncertainty and the jet energy scale uncertainty contributes 0.5\%~\cite{CMS-PAPERS-JME-10-011}.  An uncertainty of 50\% for the $\ttbar$ background contribution is due to the low event counts in the isolation distributions in high-$\ST$ bins, which are used to validate the misidentification rate. We apply a 50\% uncertainty to the normalization of all rare processes.

We define eight mutually exclusive signal regions (SRs) depending on the total number of leptons and the number of $\tauh$ candidates in the event, which are defined in Table \ref{tab:results}. Since our signal does not contain any \cPZ~bosons and does contain two to four bottom quarks, in SR1--SR4, we veto events in which any OSSF dilepton pairs have an invariant mass consistent with that of the \cPZ~boson (75--105\GeV) and require at least one $\cPqb$-tagged jet.  Each of these eight SRs is divided into five bins in $\ST$: [0--300], [300--600], [600--1000], [1000--1500], and [$>$1500]\GeV. We gain additional sensitivity in regions with $\ST>600$\GeV by removing the \cPqb-tag and \cPZ-veto requirements for events, so the SR5--SR8 contain the events that fail one or both of these requirements.

\begin{figure}[htbp]
\centering
\includegraphics[width=0.49\textwidth]{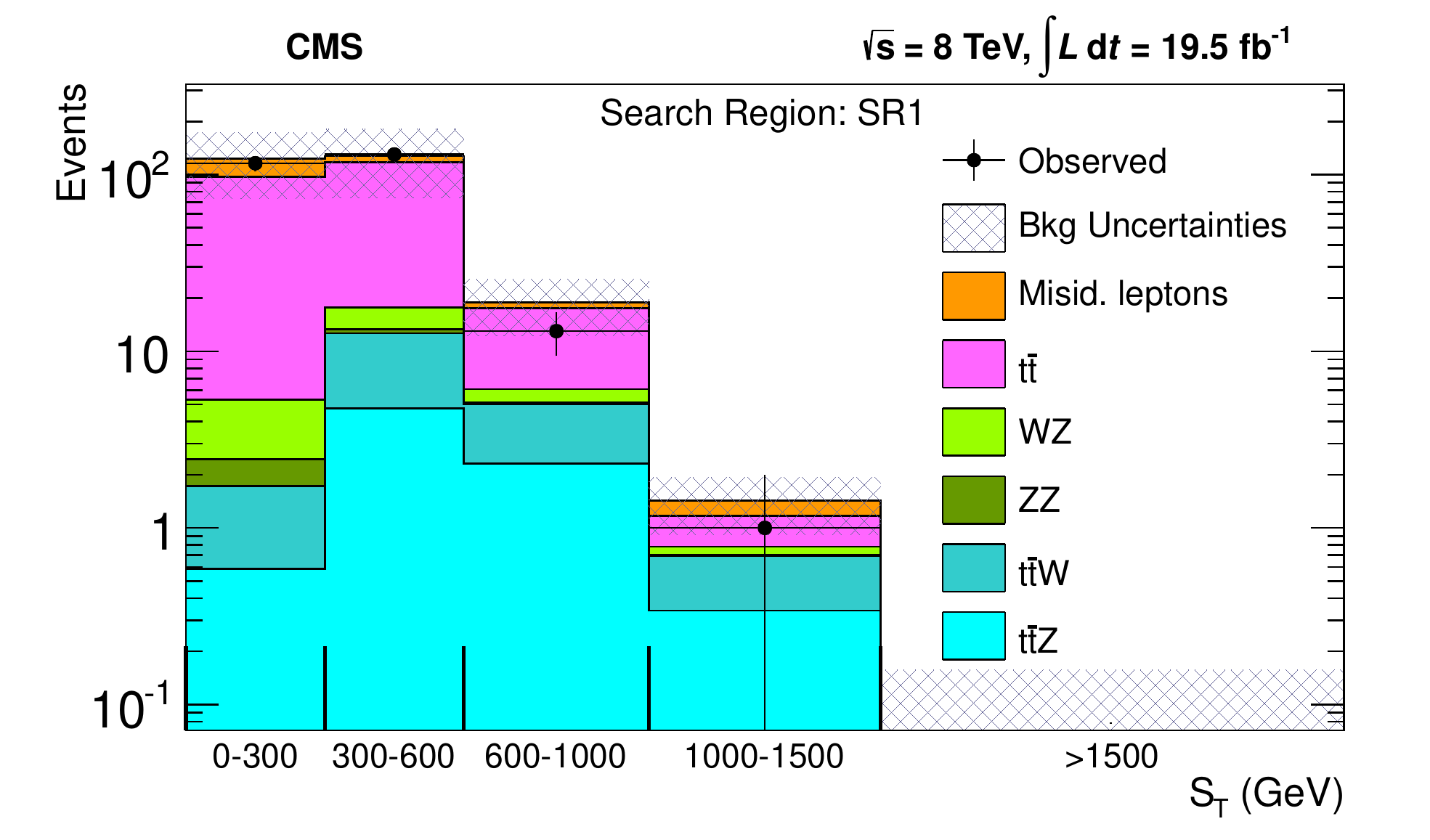}
   \caption{The $\ST$ distributions for SR1 including observed yields and background contributions. }
   \label{fig:resultsST}
\end{figure}

The observed and expected yields for SR1--SR8 are shown in Table \ref{tab:results}. We also show the $\ST$ distribution for SR1 in Fig.~\ref{fig:resultsST} with the background expectations from different sources shown separately. Data are in good agreement with the SM predictions everywhere. Please see \suppMaterial~for additional $\ST$ distributions.

\begin{table*}
\ifthenelse{\boolean{cms@external}}{\small}{\tiny}
\begin{center}
\topcaption{Observed yields for three- and four- lepton events from $\procLumi$ recorded in 2012. The channels are split by the total number of leptons (N$_\mathrm{L}$), the number of $\tauh$ candidates (N$_{\tau}$), and the $\ST$. Expected yields are the sum of simulation and estimates of backgrounds from data in each channel. SR1--SR4 require a $\cPqb$-tagged jet and veto events containing \cPZ~bosons. SR5--SR8 contain events that either contain a \cPZ~boson or have no $\cPqb$-tagged jet. The channels are mutually exclusive. The uncertainties include statistical and systematic uncertainties. The $\ST$ values are given in\GeV.
\label{tab:results}}
\begin{scotch}{ccccccccccccc}
SR & N$_\mathrm{L}$ & N$_{\tau}$ & \multicolumn{2}{c}{$0 < \ST < 300$} &\multicolumn{2}{c}{$300 < \ST < 600$}& \multicolumn{2}{c}{$600 < \ST < 1000$}& \multicolumn{2}{c}{$1000 < \ST < 1500$} & \multicolumn{2}{c}{$\ST > 1500$} \\
\hline
& & & obs & exp & obs & exp & obs & exp & obs & exp & obs & exp\\
\hline
SR1 & 3 & 0 & 116 & 123 $\pm$ 50 & 130 & 127 $\pm$ 54 & 13 & 18.9 $\pm$ 6.7 & 1 & 1.43 $\pm$ 0.51 & 0 & 0.208 $\pm$ 0.096 \\
SR2 & 3 & $\ge1$ & 710 & 698 $\pm$ 287 & 746 & 837 $\pm$ 423 & 83 & 97 $\pm$ 48 & 3 & 6.9 $\pm$ 3.9 & 0 & 0.73 $\pm$ 0.49 \\
SR3 & 4 & 0 & 0 & 0.186 $\pm$ 0.074 & 1 & 0.43 $\pm$ 0.22 & 0 & 0.19 $\pm$ 0.12 & 0 & 0.037 $\pm$ 0.039 & 0 & 0.000 $\pm$ 0.03 \\
SR4 & 4 & $\ge1$ & 1 & 0.89 $\pm$ 0.42 & 0 & 1.31 $\pm$ 0.48 & 0 & 0.39 $\pm$ 0.19 & 0 & 0.019 $\pm$ 0.026 & 0 & 0.000 $\pm$ 0.03 \\
SR5 & 3 & 0 & --- & --- & --- & --- & 152 & 161 $\pm$ 51 & 15 & 21.0 $\pm$ 8.6 & 10 & 3.45 $\pm$ 1.77 \\
SR6 &3 & 1 & --- & --- & --- & --- & 193 & 150 $\pm$ 37 & 14 & 12.8 $\pm$ 3.5 & 0 & 2.04 $\pm$ 0.79 \\
SR7&4 & 0 & --- & --- & --- & --- & 5 & 8.2 $\pm$ 2.6 & 2 & 0.93 $\pm$ 0.36 & 0 & 0.18 $\pm$ 0.08 \\
SR8&4 & 1 & --- & --- & --- & --- & 2 & 3.2 $\pm$ 0.9 & 0 & 0.28 $\pm$ 0.13 & 0 & 0.08 $\pm$ 0.05 \\
\end{scotch}
\end{center}
\end{table*}

We demonstrate natural SUSY with RPV couplings in a stop RPV model where the light stop decays to a top quark and intermediate on- or off-shell bino, $ \stone \to \widetilde{\chi}^{0*}_1 + \cPqt$. The bino decays to two leptons and a neutrino through the leptonic  RPV interactions,
$ \widetilde{\chi}^{0*}_1 \to \ell_i + \nu_j + \ell_k\quad\text{and}\quad\nu_i + \ell_j + \ell_k$,
or through the semileptonic RPV interactions,
$ \widetilde{\chi}^{0*}_1 \to \ell_i + \Pq_j + \Pq_k\quad\text{and}\quad \nu_i + \Pq_j + \Pq_k$, where the indices $i, j, k$ refer to those appearing in Eq. \ref{eqn:wrpv}. The stop is assumed to be right-handed and RPV couplings are large enough that all decays are prompt.

We generate samples to evaluate models with simplified mass spectra and leptonic RPV couplings $\Lam_{122}$ or $\Lam_{233}$. The stop masses in these samples range from 700--1250\GeV in 50\GeV steps, and bino masses range from 100--1300\GeV in 100\GeV steps. In a model with only the semi-leptonic RPV coupling $\Lamp_{233}$, we use stop masses 300--1000\GeV and bino masses  200--850\GeV, both in 50\GeV steps. In both cases, slepton and sneutrino masses are 200\GeV above the bino mass. Other particles are irrelevant in these models. Efficiency times acceptance figures for these models can be found in the \suppMaterial.

To determine which regions of phase space are excluded, we divide the channels shown in Table \ref{tab:results} by lepton flavor and perform a counting experiment using the observed event yields, the background expectations, and the signal expectations as inputs to an LHC-type CL$_\mathrm{s}$ limit calculation~\cite{ATLAS:1379837,Junk:1999kv,Read:2002hq}. A table with the finer binning is available in the \suppMaterial.

\begin{figure}[htbp]
\centering
\includegraphics[width=0.46\textwidth]{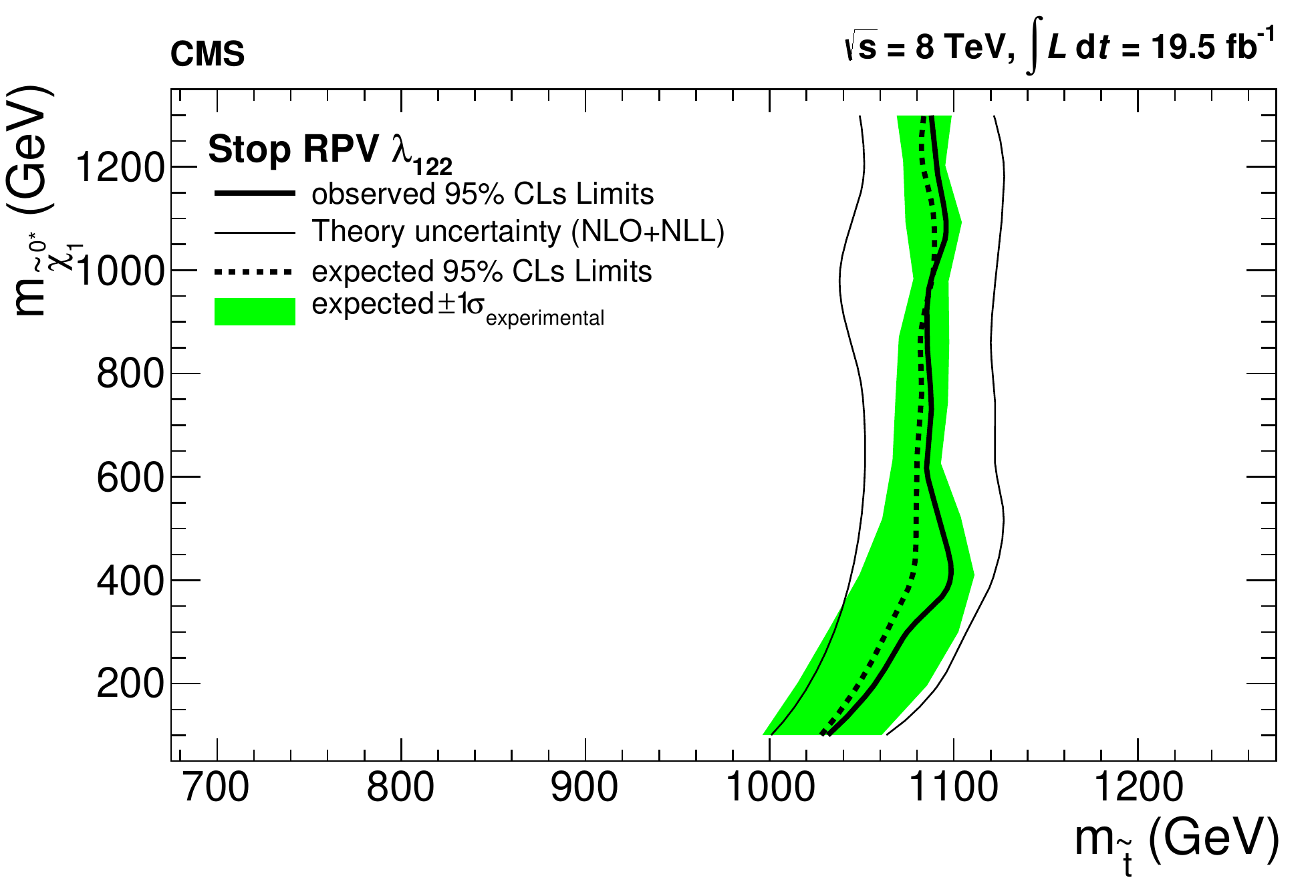}
\includegraphics[width=0.46\textwidth]{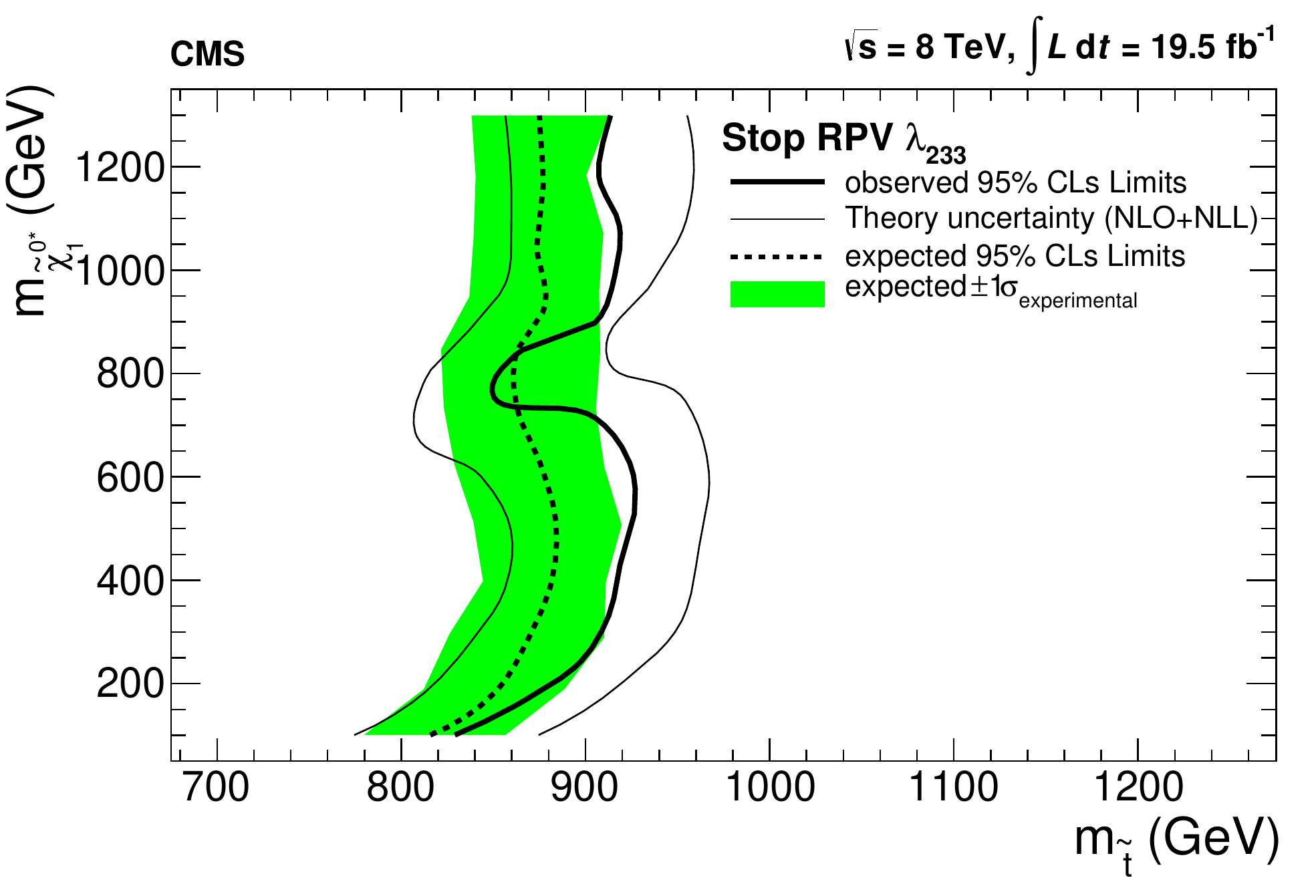}
\includegraphics[width=0.46\textwidth]{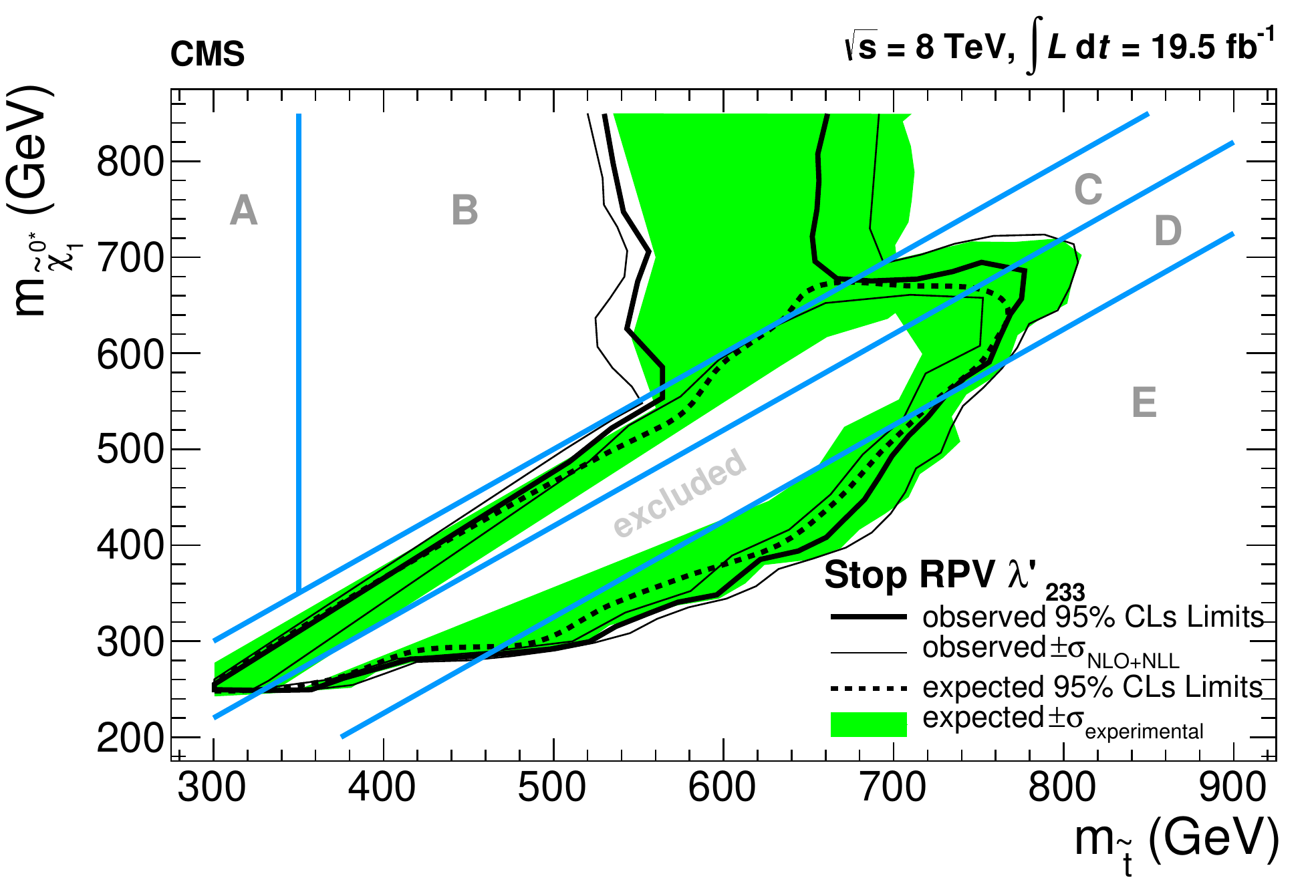}
   \caption{The 95\% confidence level limits in the stop and bino mass plane for models with RPV couplings $\Lam_{122}$, $\Lam_{233}$, and $\Lamp_{233} $. For the couplings $\Lam_{122}$ and $\Lam_{233}$, the region to the left of the curve is excluded. For $\Lamp_{233}$, the region inside the curve is excluded.
The different kinematic regions, A, B, C, D, and E, for the $\Lamp_{233}$ exclusion are explained in Table~\ref{tab:stopdecays}.
      }
   \label{fig:stopRPV}
\end{figure}

In the models with leptonic couplings, the limits are mostly independent of the bino mass, and, using the conservative minus-one-standard-deviation of the theoretical cross section with the observed result where the bino mass is 200\GeV, we exclude models with the stop mass below 1020\GeV when $\Lam_{122}$ is non-zero, and below 820\GeV when $\Lam_{233}$ is non-zero. These limits are shown in Fig. \ref{fig:stopRPV}. There is a change in kinematics at the line  $ m_{\PSGczDo} = m_{\stone} - m_{\cPqt}$, below which the stop decay is two-body, while above it is a four-body decay. Near this line, the $\PSGczDo$ and top are produced almost at rest, which results in soft leptons, reducing our acceptance. This loss of acceptance is more pronounced in the $\Lam_{233} \neq 0$ case and causes the loss of sensitivity near the line at $m_{\PSGczDo} = 800$\GeV. This feature is enhanced in the observed limit because the data has a larger statistical uncertainty in the relevant signal regions than the simulated signal samples.

In the semileptonic RPV model with $\Lamp_{233}$, there are several different kinematic regions, which are described in Table \ref{tab:stopdecays}. The most significant effect is when the decay $ \widetilde{\chi}^{0}_1 \rightarrow \mu + \cPqt + \cPqb$ is disfavored, reducing the number of leptons. The different regions where this effect is pronounced drive the shape of the exclusion for $\Lamp_{233}$. The area inside the curve is excluded. The observed limit is stronger than the expected one, which allows the observed exclusion region to reach into the regime where the bino decouples.

\begin{table}
\centering
\topcaption{\label{tab:stopdecays}
Kinematically allowed stop decay modes with RPV coupling $\Lamp_{233}$.
The allowed neutralino decay modes for $ m_{\cPqt} < m_{\PSGczDo} < m_{\stone}$
are $\PSGczDo \to \mu \cPqt \cPaqb$ and $\nu \cPqb \cPaqb$.
}
\begin{scotch}{ccc}
Label & Kinematic region & Decay mode\\
\hline
A & $ m_{\cPqt} < m_{\stone} < 2m_{\cPqt} ,  m_{\PSGczDo} $ & $\stone \rightarrow  \cPqt \nu \cPqb \cPaqb $ \\
B & $ 2 m_{\cPqt} < m_{\stone} <  m_{\PSGczDo} $ & $\stone \rightarrow \cPqt \mu \cPqt \cPaqb  $ or $ \cPqt \nu \cPqb \cPaqb $ \\
C & $ m_{\PSGczDo} < m_{\stone} < m_{\Wpm} + m_{\PSGczDo} $ & $\stone \rightarrow \ell \nu \cPqb \PSGczDo $ or $ \cmsSymbolFace{jj}\cPqb \PSGczDo $ \\
D & $m_{\Wpm} + m_{\PSGczDo} < m_{\stone} < m_{\cPqt} + m_{\PSGczDo} $ & $\stone \rightarrow \cPqb \Wpm \PSGczDo$ \\
E & $m_{\cPqt} + m_{\PSGczDo} < m_{\stone} $ & $\stone \rightarrow \cPqt \PSGczDo$ \\
\end{scotch}
\end{table}

We have performed a search for RPV supersymmetry in models with top-squark pair production using a variety of multilepton final states. Good agreement between observations and SM expectations allows us to set stringent limits on the top-squark mass in models with leptonic RPV couplings $\Lam_{122}$ and $\Lam_{233}$. For a bino mass of 200\GeV, these limits are 1020\GeV and 820\GeV, respectively. We also set limits in a model with the semi-leptonic RPV coupling $\Lamp_{233}$.

We thank Jared Evans and Yevgeny Kats for providing guidance on the signal models examined in this letter.

We congratulate our colleagues in the CERN accelerator departments for the excellent performance of the LHC and thank the technical and administrative staffs at CERN and at other CMS institutes for their contributions to the success of the CMS effort. In addition, we gratefully acknowledge the computing centres and personnel of the Worldwide LHC Computing Grid for delivering so effectively the computing infrastructure essential to our analyses. Finally, we acknowledge the enduring support for the construction and operation of the LHC and the CMS detector provided by the following funding agencies: BMWF and FWF (Austria); FNRS and FWO (Belgium); CNPq, CAPES, FAPERJ, and FAPESP (Brazil); MES (Bulgaria); CERN; CAS, MoST, and NSFC (China); COLCIENCIAS (Colombia); MSES (Croatia); RPF (Cyprus); MoER, SF0690030s09 and ERDF (Estonia); Academy of Finland, MEC, and HIP (Finland); CEA and CNRS/IN2P3 (France); BMBF, DFG, and HGF (Germany); GSRT (Greece); OTKA and NKTH (Hungary); DAE and DST (India); IPM (Iran); SFI (Ireland); INFN (Italy); NRF and WCU (Republic of Korea); LAS (Lithuania); CINVESTAV, CONACYT, SEP, and UASLP-FAI (Mexico); MSI (New Zealand); PAEC (Pakistan); MSHE and NSC (Poland); FCT (Portugal); JINR (Dubna); MON, RosAtom, RAS and RFBR (Russia); MESTD (Serbia); SEIDI and CPAN (Spain); Swiss Funding Agencies (Switzerland); NSC (Taipei); ThEPCenter, IPST, STAR and NSTDA (Thailand); TUBITAK and TAEK (Turkey); NASU (Ukraine); STFC (United Kingdom); DOE and NSF (USA).

\bibliography{auto_generated}   

\ifthenelse{\boolean{cms@external}}{}{
\clearpage
\appendix
\section{Supplementary Material\label{app:suppMat}}
To assist in the interpretation of our results in additional models, we have included the following supplemental material.

Additional \ST distributions, corresponding to SR2, SR3, and SR4, show the background breakdowns for different channels and can be found in Figs. \ref{fig:resultsSTSR2}, \ref{fig:resultsSTSR3}, and \ref{fig:resultsSTSR4}, respectively. The efficiency times acceptance distributions in Figs. \ref{fig:effaccLLE122}, \ref{fig:effaccLLE233}, and \ref{fig:effaccLQD233} show how the sensitivity of our search changes in the different kinematic regimes of the three models we investigated. Table \ref{tab:fullResults} breaks down the results shown in \ifthenelse{\boolean{cms@external}}{Table 1 in the main document}{Table~\ref{tab:results}} into much more finely binned channels, which will allow our results to be easily applied to other searches.
\begin{figure}[htb]
\centering
\includegraphics[width=0.48\textwidth]{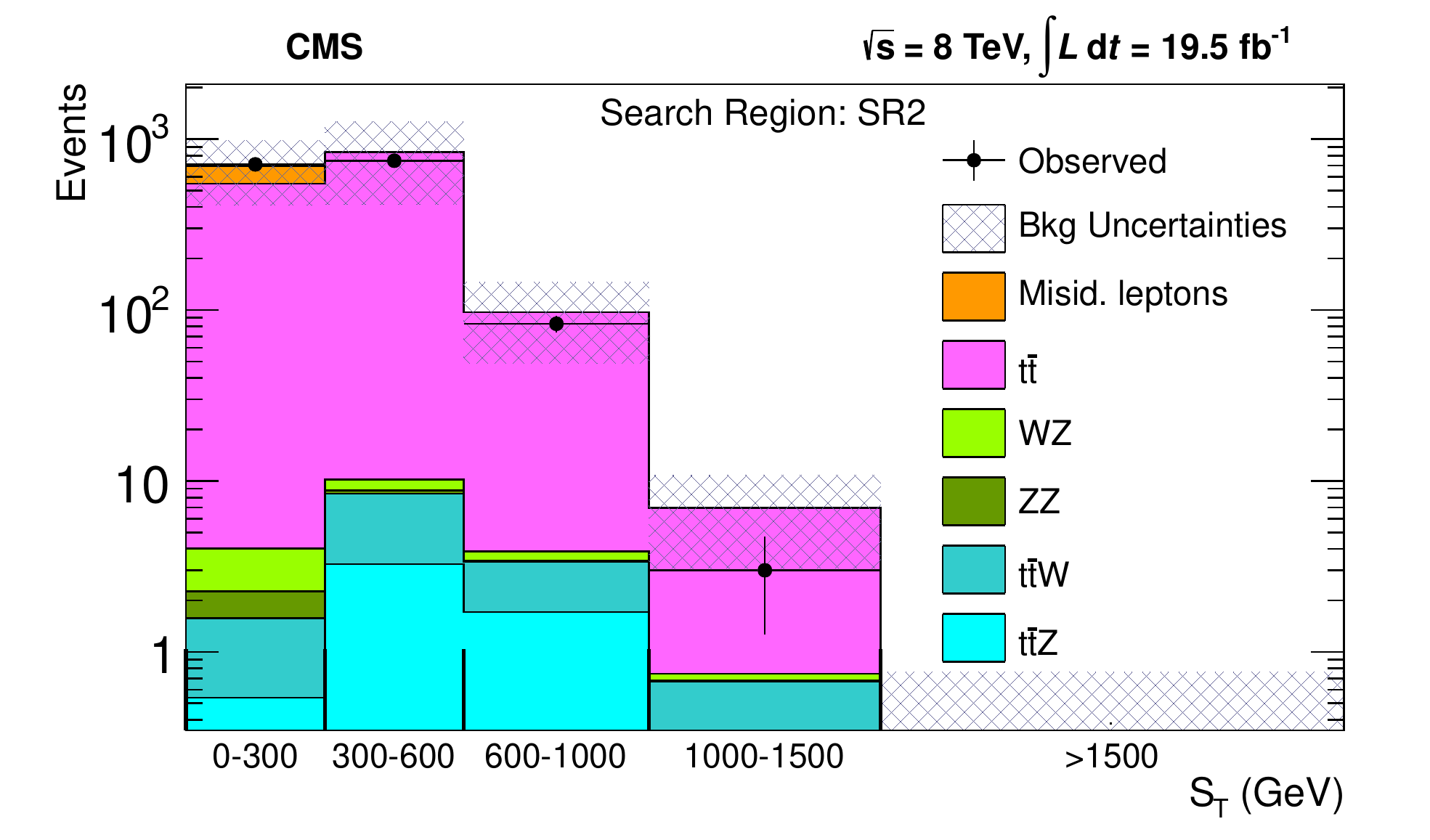}
   \caption{The $\ST$ distributions for SR2 (3 leptons including 1 \tauh) including observed yields and background contributions. }
   \label{fig:resultsSTSR2}
\end{figure}
\begin{figure}[htb]
\centering
\includegraphics[width=0.48\textwidth]{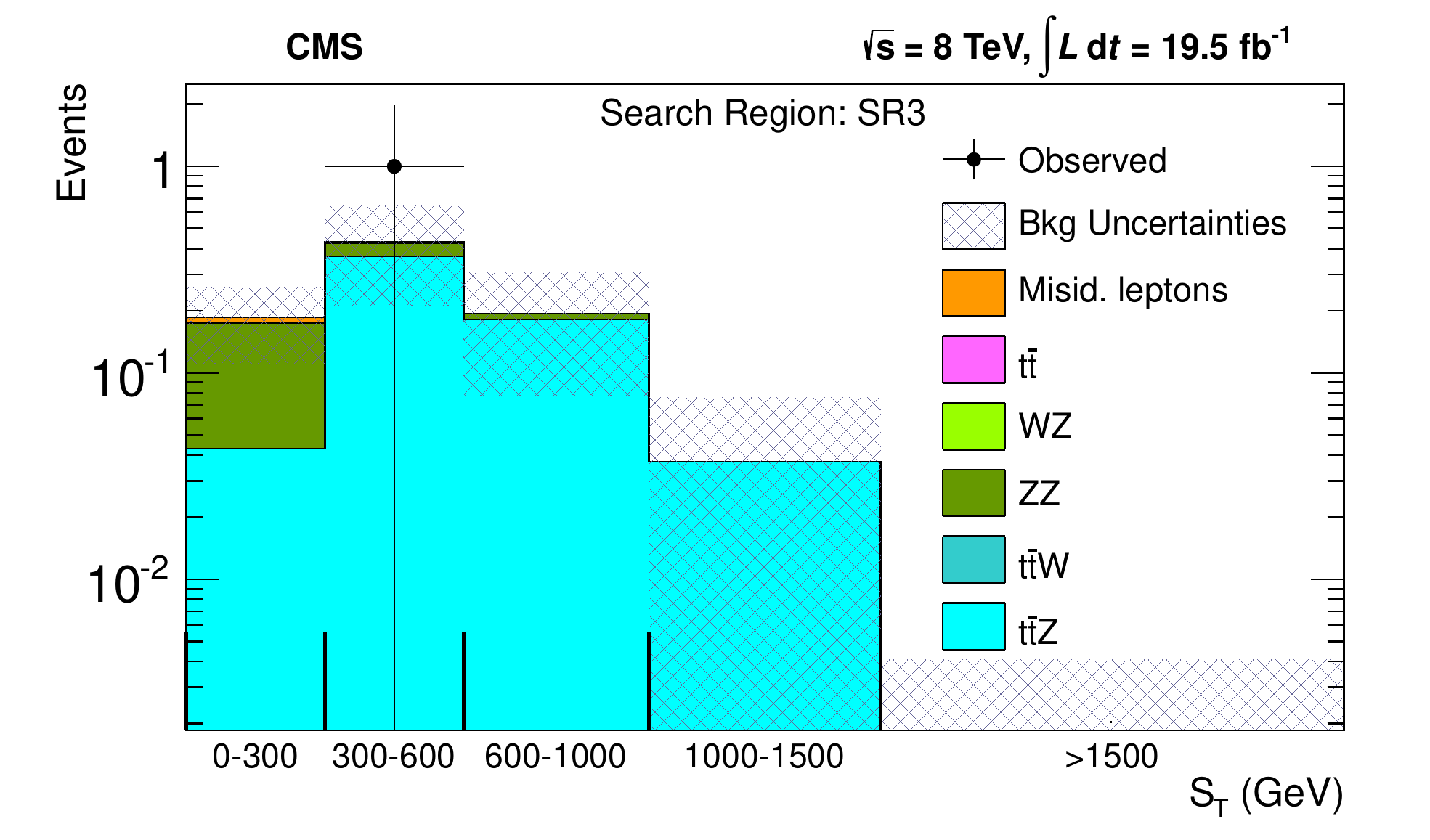}
   \caption{The $\ST$ distributions for SR3 (4 leptons including 0 \tauh) including observed yields and background contributions. }
   \label{fig:resultsSTSR3}
\end{figure}
\begin{figure}[htb]
\centering
\includegraphics[width=0.48\textwidth]{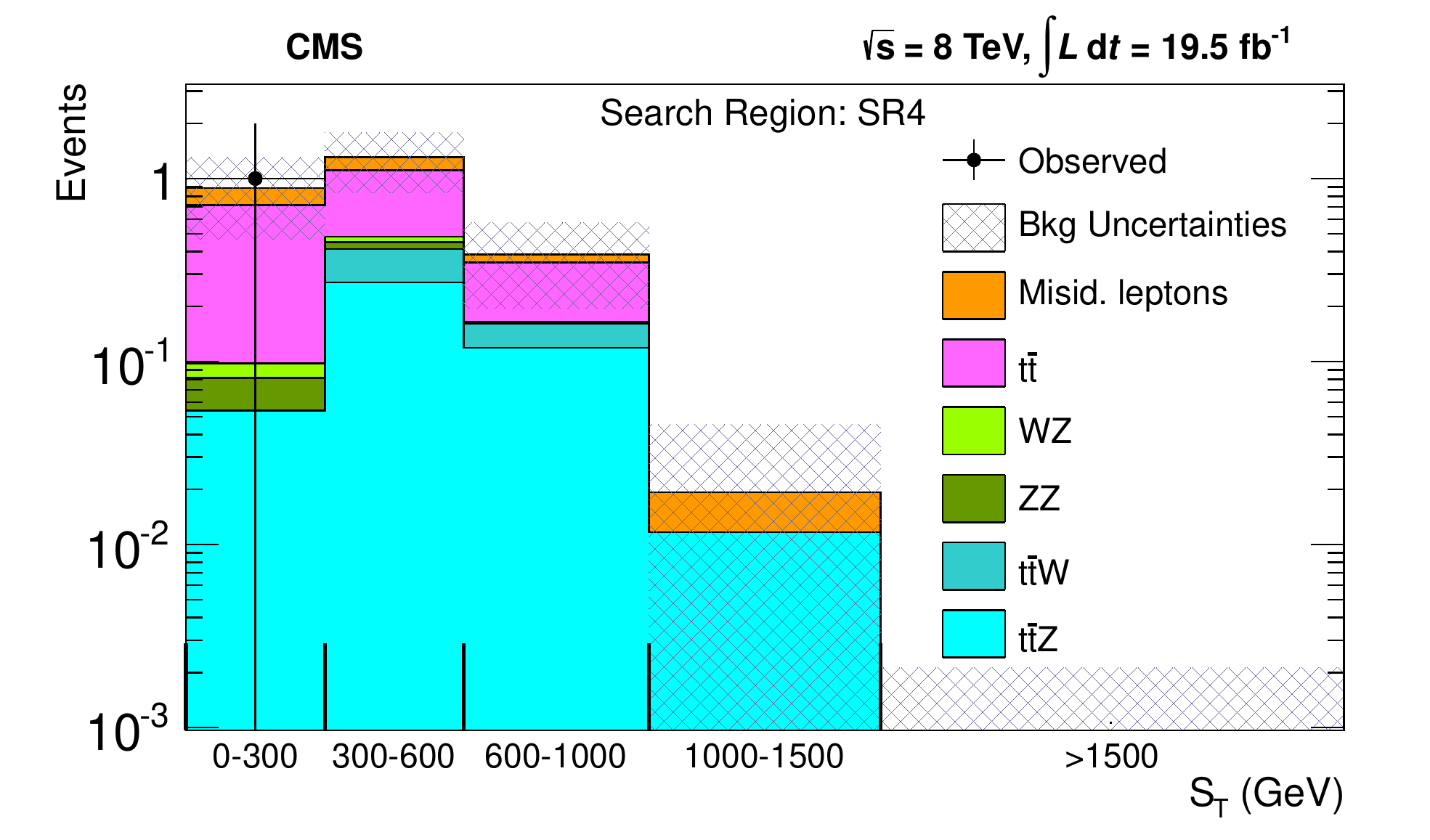}
   \caption{The $\ST$ distributions for SR4 (4 leptons including 1 \tauh) including observed yields and background contributions. }
   \label{fig:resultsSTSR4}
\end{figure}
\begin{figure}[htb]
\centering
\includegraphics[width=0.48\textwidth]{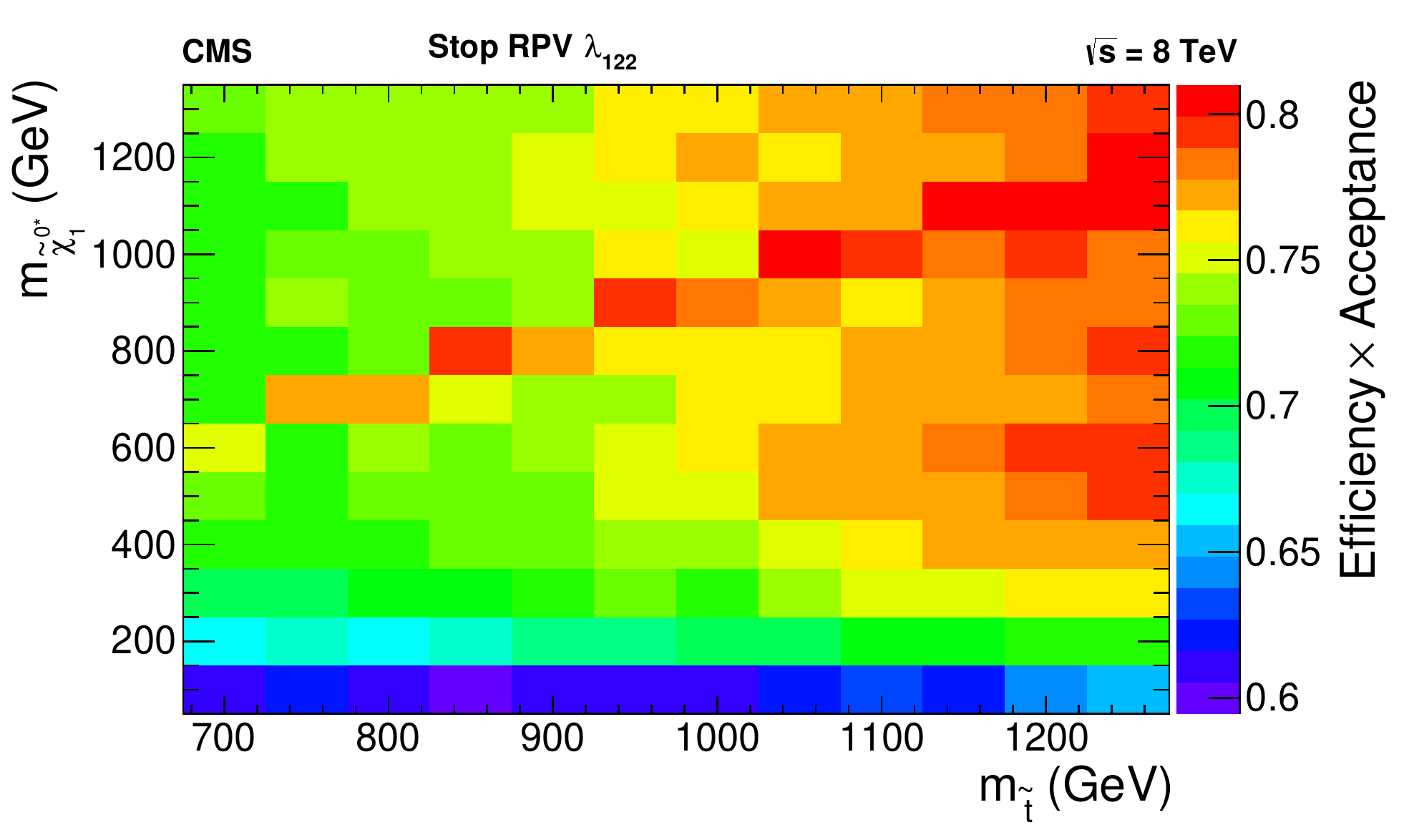}
   \caption{The efficiency times acceptance for each grid point in the stop and bino mass plane for the model with $\Lam_{122}$ non-zero.}
   \label{fig:effaccLLE122}
\end{figure}
\begin{figure}[htb]
\centering
\includegraphics[width=0.48\textwidth]{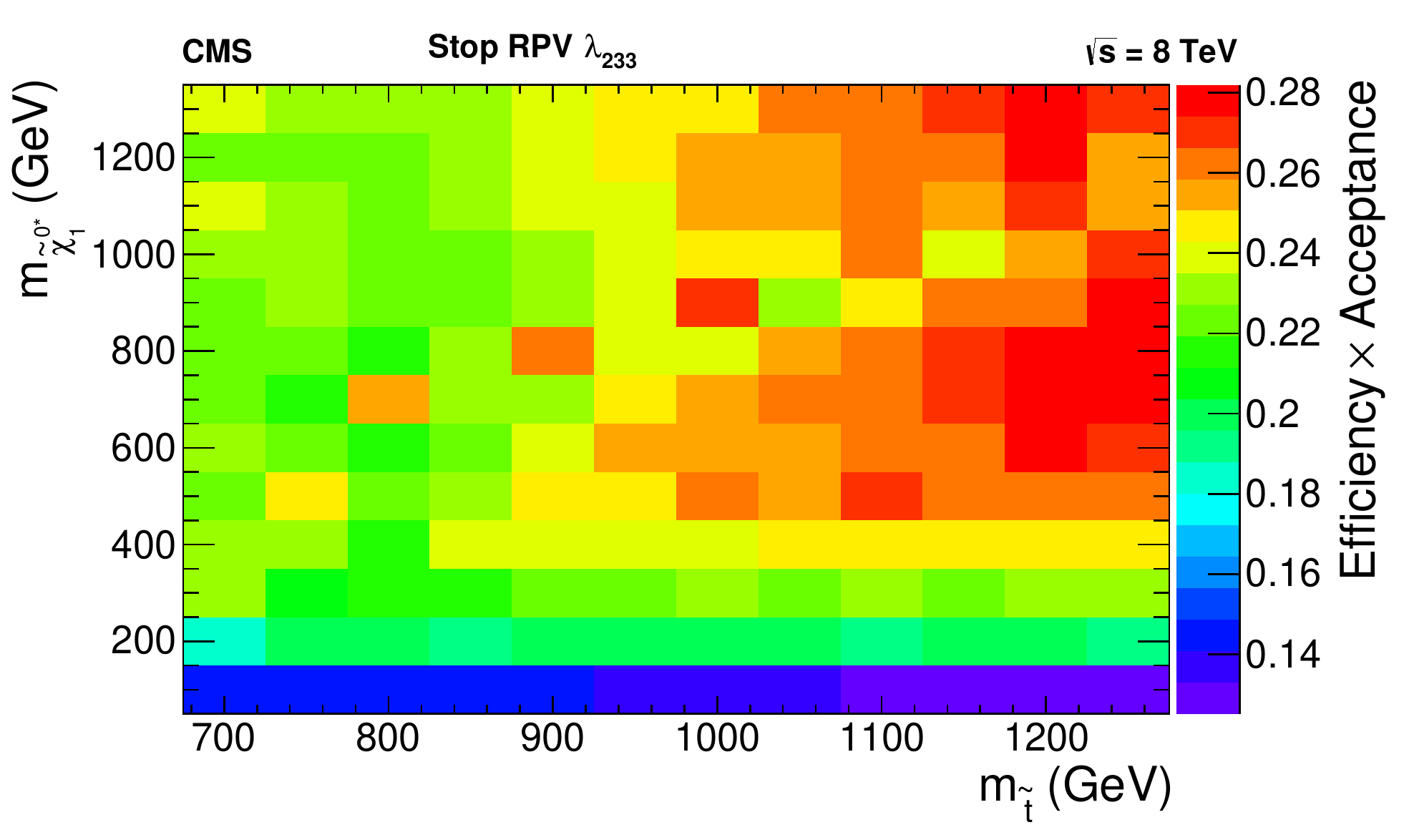}
   \caption{The efficiency times acceptance for each grid point in the stop and bino mass plane for the model with $\Lam_{233}$ non-zero. }
   \label{fig:effaccLLE233}
\end{figure}
\begin{figure}[htb]
\centering
\includegraphics[width=0.48\textwidth]{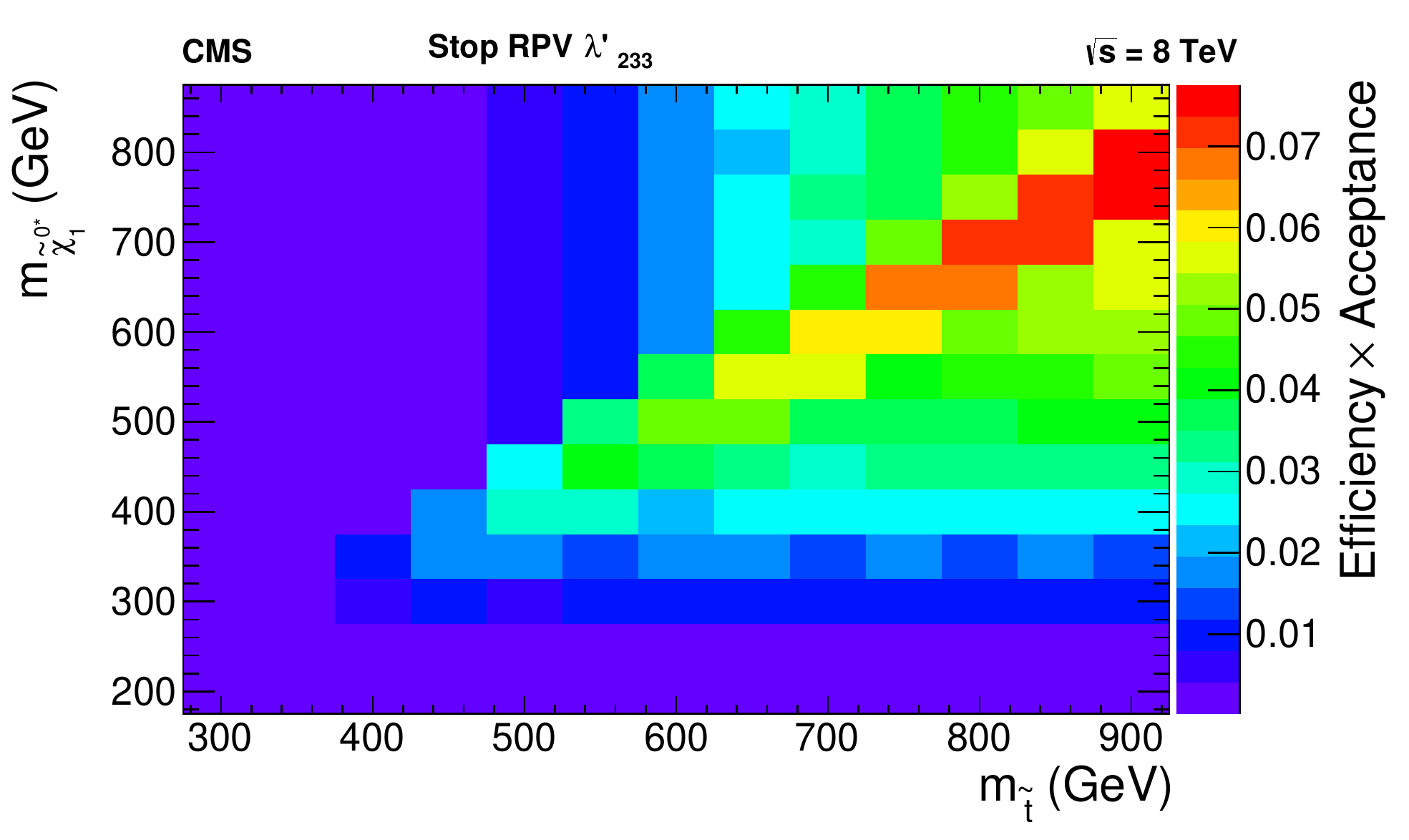}
   \caption{The efficiency times acceptance for each grid point in the stop and bino mass plane for the model with $\Lamp_{233}$ non-zero.}
   \label{fig:effaccLQD233}
\end{figure}
\begin{table*}[htb]
\tiny
\centering
\topcaption{Observed yields for three- and four- lepton events from $\procLumi$ recorded in 2012. The channels are split by the total number of leptons (N$_\mathrm{L}$), the number of $\tauh$ candidates (N$_{\tau}$), whether the event contains $\cPqb$-tagged jets (N$_{\cPqb}$), the number of OSSF pairs (N$_{\textrm OSSF}$), binning in the dilepton invariant mass ($m_{\ell\ell}$), and the $\ST$. Expected yields are the sum of simulation and estimates of backgrounds from data in each channel.  The channels are mutually exclusive. The uncertainties include statistical and systematic uncertainties. The $\ST$ values are given in\GeV.
\label{tab:fullResults}}
\resizebox{\textwidth}{!}{
\begin{scotch}{ccccccccccccccc}
N$_\mathrm{L}$ & N$_{\tau}$ & N$_{\cPqb}$ & N$_{\textrm OSSF}$ & $m_{\ell\ell}$& \multicolumn{2}{c}{$0 < \ST < 300$} &\multicolumn{2}{c}{$300 < \ST < 600$}& \multicolumn{2}{c}{$600 < \ST < 1000$}& \multicolumn{2}{c}{$1000 < \ST < 1500$} & \multicolumn{2}{c}{$\ST > 1500$} \\
\hline
& & & & & obs & exp & obs & exp & obs & exp & obs & exp & obs & exp\\
\hline
4 & 0 & 0 & 0 & -- &0 & 0.06 $\pm$ 0.06 & 0 & 0.09 $\pm$ 0.07 & 0 & 0.00 $\pm$ 0.03 & 0 & 0.00 $\pm$ 0.03 & 0 & 0.00 $\pm$ 0.03 \\
4 & 0 & 1 & 0 & -- &0 & 0.00 $\pm$ 0.03 & 0 & 0.00 $\pm$ 0.03 & 0 & 0.06 $\pm$ 0.05 & 0 & 0.00 $\pm$ 0.03 & 0 & 0.00 $\pm$ 0.03 \\
4 & 0 & 0 & 1 & on-Z & 2 & 3.1 $\pm$ 0.90 & 5 & 1.9 $\pm$ 0.48 & 0 & 0.44 $\pm$ 0.16 & 1 & 0.06 $\pm$ 0.06 & 0 & 0.00 $\pm$ 0.03 \\
4 & 0 & 1 & 1 & on-Z & 2 & 0.07 $\pm$ 0.05 & 2 & 1.1 $\pm$ 0.53 & 0 & 0.57 $\pm$ 0.30 & 0 & 0.12 $\pm$ 0.09 & 0 & 0.02 $\pm$ 0.03 \\
4 & 0 & 0 & 1 & off-Z & 2 & 0.48 $\pm$ 0.18 & 0 & 0.27 $\pm$ 0.11 & 0 & 0.07 $\pm$ 0.05 & 0 & 0.00 $\pm$ 0.02 & 0 & 0.00 $\pm$ 0.03 \\
4 & 0 & 1 & 1 & off-Z & 0 & 0.04 $\pm$ 0.04 & 0 & 0.34 $\pm$ 0.17 & 0 & 0.06 $\pm$ 0.08 & 0 & 0.04 $\pm$ 0.04 & 0 & 0.00 $\pm$ 0.03 \\
4 & 0 & 0 & 2 & on-Z & 135 & 120 $\pm$ 29 & 26 & 43 $\pm$ 10 & 4 & 6.0 $\pm$ 2.0 & 1 & 0.63 $\pm$ 0.26 & 0 & 0.06 $\pm$ 0.04 \\
4 & 0 & 1 & 2 & on-Z & 1 & 1.0 $\pm$ 0.27 & 4 & 3.2 $\pm$ 1.1 & 1 & 1.1 $\pm$ 0.39 & 0 & 0.11 $\pm$ 0.06 & 0 & 0.04 $\pm$ 0.04 \\
4 & 0 & 0 & 2 & off-Z & 7 & 8.3 $\pm$ 2.3 & 3 & 1.1 $\pm$ 0.30 & 0 & 0.11 $\pm$ 0.05 & 0 & 0.01 $\pm$ 0.02 & 0 & 0.00 $\pm$ 0.02 \\
4 & 0 & 1 & 2 & off-Z & 0 & 0.18 $\pm$ 0.07 & 1 & 0.22 $\pm$ 0.11 & 0 & 0.15 $\pm$ 0.08 & 0 & 0.00 $\pm$ 0.03 & 0 & 0.00 $\pm$ 0.03 \\
4 & 1 & 0 & 0 & -- &2 & 1.1 $\pm$ 0.46 & 1 & 0.54 $\pm$ 0.20 & 0 & 0.12 $\pm$ 0.12 & 0 & 0.00 $\pm$ 0.03 & 0 & 0.00 $\pm$ 0.03 \\
4 & 1 & 1 & 0 & -- &0 & 0.26 $\pm$ 0.16 & 0 & 0.29 $\pm$ 0.13 & 0 & 0.13 $\pm$ 0.11 & 0 & 0.01 $\pm$ 0.02 & 0 & 0.00 $\pm$ 0.03 \\
4 & 1 & 0 & 1 & on-Z & 43 & 42 $\pm$ 11 & 10 & 12 $\pm$ 3.1 & 0 & 1.8 $\pm$ 0.63 & 0 & 0.11 $\pm$ 0.07 & 0 & 0.02 $\pm$ 0.03 \\
4 & 1 & 1 & 1 & on-Z & 2 & 1.0 $\pm$ 0.40 & 2 & 1.7 $\pm$ 0.5 & 0 & 0.78 $\pm$ 0.33 & 0 & 0.04 $\pm$ 0.04 & 0 & 0.01 $\pm$ 0.03 \\
4 & 1 & 0 & 1 & off-Z & 18 & 8.4 $\pm$ 2.2 & 4 & 2.1 $\pm$ 0.52 & 2 & 0.48 $\pm$ 0.18 & 0 & 0.13 $\pm$ 0.08 & 0 & 0.01 $\pm$ 0.03 \\
4 & 1 & 1 & 1 & off-Z & 1 & 0.64 $\pm$ 0.31 & 0 & 1.2 $\pm$ 0.44 & 0 & 0.30 $\pm$ 0.13 & 0 & 0.02 $\pm$ 0.03 & 0 & 0.00 $\pm$ 0.03 \\
3 & 0 & 0 & 0 & -- &72 & 80 $\pm$ 23 & 32 & 27 $\pm$ 11 & 3 & 3.1 $\pm$ 1.00 & 0 & 0.22 $\pm$ 0.18 & 0 & 0.07 $\pm$ 0.06 \\
3 & 0 & 1 & 0 & -- &37 & 33 $\pm$ 16 & 42 & 39 $\pm$ 19 & 2 & 5.0 $\pm$ 2.0 & 0 & 0.36 $\pm$ 0.14 & 0 & 0.06 $\pm$ 0.07 \\
3 & 0 & 0 & 1 & on-Z & 4255 & 4400 $\pm$ 690 & 669 & 740 $\pm$ 170 & 106 & 110 $\pm$ 41 & 11 & 15 $\pm$ 6.9 & 3 & 1.3 $\pm$ 0.76 \\
3 & 0 & 1 & 1 & on-Z & 140 & 150 $\pm$ 25 & 122 & 110 $\pm$ 25 & 16 & 25 $\pm$ 7.0 & 2 & 3.3 $\pm$ 1.2 & 1 & 0.32 $\pm$ 0.22 \\
3 & 0 & 0 & 1 & $m_{\ell\ell} < 75$ (\GeVns{}) & 617 & 640 $\pm$ 100 & 84 & 86 $\pm$ 21 & 14 & 11 $\pm$ 3.6 & 0 & 1.2 $\pm$ 0.39 & 1 & 0.12 $\pm$ 0.09 \\
3 & 0 & 1 & 1 & $m_{\ell\ell} < 75$ (\GeVns{}) & 62 & 74 $\pm$ 28 & 52 & 57 $\pm$ 23 & 4 & 8.3 $\pm$ 2.7 & 1 & 0.69 $\pm$ 0.28 & 0 & 0.08 $\pm$ 0.06 \\
3 & 0 & 0 & 1 & $m_{\ell\ell} > 105$ (\GeVns{}) & 180 & 200 $\pm$ 34 & 63 & 66 $\pm$ 12 & 13 & 10 $\pm$ 2.5 & 2 & 1.1 $\pm$ 0.40 & 0 & 0.16 $\pm$ 0.09 \\
3 & 0 & 1 & 1 & $m_{\ell\ell} > 105$ (\GeVns{}) & 17 & 17 $\pm$ 6.5 & 36 & 35 $\pm$ 14 & 7 & 7.4 $\pm$ 2.5 & 0 & 0.54 $\pm$ 0.23 & 0 & 0.08 $\pm$ 0.05 \\
3 & 1 & 0 & 0 & -- &1194 & 1300 $\pm$ 330 & 289 & 290 $\pm$ 130 & 26 & 28 $\pm$ 12 & 2 & 2.6 $\pm$ 1.3 & 0 & 0.23 $\pm$ 0.20 \\
3 & 1 & 1 & 0 & -- &316 & 330 $\pm$ 160 & 410 & 480 $\pm$ 240 & 46 & 58 $\pm$ 28 & 2 & 3.9 $\pm$ 2.0 & 0 & 0.46 $\pm$ 0.32 \\
3 & 1 & 0 & 1 & on-Z & 49916 & 49000 $\pm$ 15000 & 2099 & 2700 $\pm$ 770 & 108 & 70 $\pm$ 17 & 9 & 6.0 $\pm$ 1.6 & 0 & 0.33 $\pm$ 0.18 \\
3 & 1 & 1 & 1 & on-Z & 795 & 830 $\pm$ 230 & 325 & 280 $\pm$ 74 & 17 & 17 $\pm$ 4.8 & 1 & 1.8 $\pm$ 0.64 & 0 & 0.30 $\pm$ 0.14 \\
3 & 1 & 0 & 1 & $m_{\ell\ell} < 75$ (\GeVns{}) & 10173 & 9200 $\pm$ 2700 & 290 & 280 $\pm$ 72 & 21 & 11 $\pm$ 3.5 & 1 & 0.97 $\pm$ 0.44 & 0 & 0.04 $\pm$ 0.06 \\
3 & 1 & 1 & 1 & $m_{\ell\ell} < 75$ (\GeVns{}) & 297 & 290 $\pm$ 97 & 167 & 170 $\pm$ 87 & 14 & 12 $\pm$ 6.0 & 0 & 1.1 $\pm$ 0.74 & 0 & 0.06 $\pm$ 0.08 \\
3 & 1 & 0 & 1 & $m_{\ell\ell} > 105$ (\GeVns{}) & 1620 & 1700 $\pm$ 480 & 285 & 370 $\pm$ 96 & 21 & 23 $\pm$ 7.2 & 1 & 1.4 $\pm$ 0.61 & 0 & 0.22 $\pm$ 0.23 \\
3 & 1 & 1 & 1 & $m_{\ell\ell} > 105$ (GeV) & 97 & 79 $\pm$ 36 & 169 & 190 $\pm$ 94 & 23 & 28 $\pm$ 14 & 1 & 2.2 $\pm$ 1.3 & 0 & 0.20 $\pm$ 0.18\\ 
\end{scotch}}
\end{table*}

}\cleardoublepage \appendix\section{The CMS Collaboration \label{app:collab}}\begin{sloppypar}\hyphenpenalty=5000\widowpenalty=500\clubpenalty=5000\textbf{Yerevan Physics Institute,  Yerevan,  Armenia}\\*[0pt]
S.~Chatrchyan, V.~Khachatryan, A.M.~Sirunyan, A.~Tumasyan
\vskip\cmsinstskip
\textbf{Institut f\"{u}r Hochenergiephysik der OeAW,  Wien,  Austria}\\*[0pt]
W.~Adam, T.~Bergauer, M.~Dragicevic, J.~Er\"{o}, C.~Fabjan\cmsAuthorMark{1}, M.~Friedl, R.~Fr\"{u}hwirth\cmsAuthorMark{1}, V.M.~Ghete, N.~H\"{o}rmann, J.~Hrubec, M.~Jeitler\cmsAuthorMark{1}, W.~Kiesenhofer, V.~Kn\"{u}nz, M.~Krammer\cmsAuthorMark{1}, I.~Kr\"{a}tschmer, D.~Liko, I.~Mikulec, D.~Rabady\cmsAuthorMark{2}, B.~Rahbaran, C.~Rohringer, H.~Rohringer, R.~Sch\"{o}fbeck, J.~Strauss, A.~Taurok, W.~Treberer-Treberspurg, W.~Waltenberger, C.-E.~Wulz\cmsAuthorMark{1}
% ...
\vskip\cmsinstskip
% ...
1:~~Also at Vienna University of Technology, Vienna, Austria\\
2:~~Also at CERN, European Organization for Nuclear Research, Geneva, Switzerland\\
% ...

\end{sloppypar}
\end{document}